\documentclass[12pt]{article}

\begin{document}

\author{E. Ahmed$^{1,2}$, A. S. Hegazi$^{1,2}$ and A. S. Elgazzar$^{3}$ \\
$^{1}$Mathematics Department, Faculty of Science\\
35516 Mansoura, Egypt\\
$^{2}$Mathematics Department, Faculty of Science\\
Al-Ain PO Box 17551, UAE\\
$^{3}$Mathematics Department, Faculty of Education\\
45111 El-Arish, Egypt}
\title{On Spatial Asymmetric Games}
\date{}
\maketitle

\begin{abstract}
The stability of some spatial asymmetric games is discussed. Both
linear and nonlinear asymptotic stability of asymmetric hawk-dove
and prisoner's dilemma are studied. Telegraph reaction diffusion
equations for the asymmetric spatial games are presented.
Asymmetric game of parental investment is studied in the presence
of both ordinary and cross diffusions.
\end{abstract}

\section{Introduction}

In asymmetric games [Hofbauer and Sigmund 1998], different players have
different strategies and different payoffs. In reality, most games are
asymmetric e.g. battle of the sexes and owners-intruders games. The
differential equations of the asymmetrical games are

\begin{equation}
\begin{array}{c}
\frac{\mathrm{d}p_{i}}{\mathrm{d}t}=p_{i}\left[ (Aq)_{i}-pAq\right] \\
\frac{\mathrm{d}q_{i}}{\mathrm{d}t}=q_{i}\left[ (Bp)_{i}-qBp\right]
\end{array}
,\;i=1,2,...,n,
\end{equation}
where $A(B)$ is the payoff matrix of the first (second) player, and $%
p_{i}(q_{i})$ is the fraction of adopters of the strategy $i$ in the first
(second) population, respectively.

The problem of Turing instability (diffusion induced instability) [Okubo
1980] for symmetric games have been already studied [Cressman and Vickers
1997]. In this case, the standard replicator\ equation for the symmetric
game is given by

\begin{equation}
\frac{\partial p_{i}}{\partial t}=p_{i}\left[ (Ap)_{i}-pAp\right] +D\nabla
^{2}p_{i}
\end{equation}

It is known [Okubo 1980] that including spatial effects may significantly
change the stability of equilibrium points. Also spatial effects are crucial
in many biological phenomena. Some authors [Boerlijst and Hogeweg 1995] have
studied general spatial hypercycle systems. Therefore studying spatial
asymmetric games is an important problem. Due to the difficulty in defining
evolutionarily stable strategy (ESS) in asymmetric games [Hofbauer and
Sigmund 1998], we will concentrate on asymptotically stable strategies. The
equations of spatial asymmetric games are [Hofbauer et al 1997]

\begin{equation}
\begin{array}{c}
\frac{\partial p_{i}}{\partial t}=p_{i}\left[ (Aq)_{i}-pAq\right]
+D_{1}\nabla ^{2}p_{i}, \\
\frac{\partial q_{i}}{\partial t}=q_{i}\left[ (Bp)_{i}-qBp\right]
+D_{2}\nabla ^{2}q_{i}.
\end{array}
\end{equation}

In this paper, we will attempt to answer the following questions:

\begin{enumerate}
\item  Given an asymptotically stable solution to the system (1), does
Turing instability exist for the corresponding spatial game (3)?

\item  Given an asymptotically unstable solution to the system (1), can
diffusion stabilize it?

\item  Given an asymptotically linearly stable solution to the system (1),
is it nonlinearly stable?
\end{enumerate}

Our typical examples will be the asymmetric hawk-dove (AHD) and the
asymmetric prisoner's dilemma (APD) games.

The paper is organized as follows: In section 2, the asymmetric hawk-dove
game is studied. Conditions for Turing stability and nonlinear finite
amplitude instability are derived. In section 3, The asymmetric prisoner's
dilemma is presented. Telegraph reaction diffusion equation is applied for
stability analysis of the asymmetric prisoner's dilemma game in section 4.
In section 5, an asymmetric game of parental investment will be studied.
Some conclusions are summarized in section 6.

\section{Asymmetric hawk-dove game}

In this case the possible strategies are hawk (H) and dove (D), and the
payoff matrices $A$ and $B$ in Eq. (1) become

\begin{equation}
A=\left[
\begin{array}{cc}
\frac{1}{2}(v_{1}-c_{1}) & v_{1} \\
0 & \frac{v_{1}}{2}
\end{array}
\right] ,\;B=\left[
\begin{array}{cc}
\frac{1}{2}(v_{2}-c_{2}) & v_{2} \\
0 & \frac{v_{2}}{2}
\end{array}
\right] ,
\end{equation}
where $c_{i}>v_{i},\;i=1,2.$ The corresponding partial differential
equations for the spatial AHD are

\begin{equation}
\begin{array}{c}
\frac{\partial p}{\partial t}=D_{1}\nabla ^{2}p\;+\frac{1}{2}%
p(1-p)(v_{1}-c_{1}q), \\
\frac{\partial q}{\partial t}=D_{2}\nabla ^{2}q\;+\frac{1}{2}%
q(1-q)(v_{2}-c_{2}p),
\end{array}
\end{equation}
where $p\;(q)$ is the fraction of hawks in the population of the first
(second) player. It is direct to see that the solution $p=1,\;q=0$ is
linearly asymptotically stable solution for the system (5), without
diffusion ($D_{1}=$ $D_{2}=0$).

The first question is about Turing (diffusion induced) instability [Okubo
1980]. It occurs if the following system

\begin{equation}
\begin{array}{c}
\frac{\partial p}{\partial t}=D_{1}\frac{\partial ^{2}p}{\partial x^{2}}%
+f_{1}(p,q), \\
\frac{\partial q}{\partial t}=D_{2}\frac{\partial ^{2}q}{\partial x^{2}}%
+f_{2}(p,q),
\end{array}
\end{equation}
has an equilibrium solution $(p_{ss},q_{ss})\;$which is stable if $D_{1}=$ $%
D_{2}=0,$ and the corresponding linearized system

\begin{equation}
\begin{array}{c}
p=p_{ss}+\varepsilon (x,t),\;q=q_{ss}+\eta (x,t), \\
\frac{\partial \varepsilon }{\partial t}=D_{1}\frac{\partial ^{2}\varepsilon
}{\partial x^{2}}+a_{11}\varepsilon +a_{12}\eta , \\
\frac{\partial \eta }{\partial t}=D_{2}\frac{\partial ^{2}\eta }{\partial
x^{2}}+a_{21}\varepsilon +a_{22}\eta ,
\end{array}
\end{equation}
satisfies the condition

\begin{equation}
H(k^{2})=D_{1}D_{2}k^{4}-(D_{1}a_{22}+D_{2}a_{11})k^{2}+a_{11}a_{22}-a_{12}a_{21}<0
\end{equation}
In this case diffusion will destabilize the solution $(p_{ss},q_{ss}).$ This
is Turing instability.

Applying the above procedure to the spatial AHD one gets:\newline
\newline
\textbf{Proposition (1):} The equilibrium solution $p=1,\;q=0$ of the AHD
game is Turing stable.\newline

The second question to be discussed is: Can diffusion stabilize an unstable
solution of the AHD game? Consider the internal solution $%
p=v_{2}/c_{2},\;q=v_{1}/c_{1}$. It is asymptotically unstable solution to
the diffusionless case ($D_{1}=$ $D_{2}=0$). Including diffusion and
linearizing around $p=v_{2}/c_{2},\;q=v_{1}/c_{1}$, \ and assuming the
following boundary conditions:

\begin{equation}
\begin{array}{c}
p=\frac{v_{2}}{c_{2}}+\varepsilon (x,t),\;q=\frac{v_{1}}{c_{1}}+\eta (x,t),
\\
\varepsilon (0,t)=\varepsilon (1,t)=0,\;\eta (0,t)=\eta (1,t)=0,
\end{array}
\end{equation}
then\newline
\newline
\textbf{Proposition (2):} The interior solution $p=v_{2}/c_{2},%
\;q=v_{1}/c_{1}$, with the boundary conditions (9) is linearly
asymptotically stable if

\begin{equation}
D_{1}D_{2}\pi ^{4}-\frac{v_{1}v_{2}}{4}(1-\frac{v_{1}}{c_{1}})(1-\frac{v_{2}%
}{c_{2}})\geq 0.
\end{equation}
\newline

Linear stability analysis is useful if the perturbations of equilibrium are
infinitesimally small. This is not always the case in biological systems. In
this case one has to study finite amplitude instability (FAI) [Stuart 1989]
of the equilibrium solution. In the following, we generalize the work of
Stuart to the two species case. Consider the following equation

\begin{equation}
\frac{\partial \theta }{\partial t}=\frac{\partial ^{2}\theta }{\partial
x^{2}}+f(\theta ),\;\theta (0,t)=\theta (1,t)=0.
\end{equation}
Linearizing around the solution\ $\theta =0$, i.e.\ let

\[
\theta (x,t)=v(x,t),
\]
linearize in $v$, then

\begin{equation}
\begin{array}{c}
\frac{\partial v}{\partial t}=\frac{\partial ^{2}v}{\partial x^{2}}%
+vf^{\prime }(0),\;\mathrm{Let}\;v=\phi (x)\;\exp (\sigma t)\Rightarrow \\
\sigma \phi =\phi ^{\prime \prime }+f^{\prime }(0)\phi ,\;\phi (0)=\phi
(1)=0.
\end{array}
\end{equation}
Set $\phi =\sum_{l}a_{l}\sin (\pi lx),$ then bifurcation points are given by
$f^{\prime }(0)=(\pi l)^{2},\;l=1,2,\ldots $. Studying the stability of the
first bifurcation point $l=1$ using Matkowsky two-time nonlinear stability
analysis, one defines $\lambda =a-k$,\ set$\;\lambda =\pi ^{2}+\lambda
_{0}\varepsilon ^{2},\;\varepsilon $ is a small parameter. Decompose the
time into fast $t^{\prime }$ and slow $\tau $, then

\[
\frac{\partial }{\partial t}=\frac{\partial }{\partial t^{\prime }}%
+\varepsilon ^{2}\frac{\partial }{\partial \tau }.
\]
Expand $u$ in powers of $\varepsilon ^{2}$, then$\;u\simeq \varepsilon
v_{1}+\varepsilon ^{3}v_{3}+\ldots $, (notice that $f^{\prime \prime }(0)=0$%
), then substituting in Eq. (12), one gets

\[
v_{1}(x,t^{\prime },\tau )=\sum_{l}a_{l}(\tau )\sin (\pi lx)\exp
(1-l^{2})\pi ^{2}t^{\prime }.
\]
Substituting into the cubic term and setting the constant term in $t^{\prime
}$ equal to zero, one finally gets

\[
\begin{array}{c}
\frac{da_{1}}{d\tau }=\lambda _{0}a_{1}+b(a_{1})^{3}, \\
b=4\pi ^{4}f^{\prime }(0)\frac{\int_{0}^{1}\sin ^{4}\pi x\;\mathrm{d}x}{%
6\int_{0}^{1}\sin ^{2}\pi x\;\mathrm{d}x}.
\end{array}
\]
Thus a nonlinear (finite amplitude) instability arises if

\begin{equation}
\lambda _{0}<0\;\mathrm{and}\;\left| a_{1}(0)\right| >\frac{\sqrt{\left|
\lambda _{0}\right| }}{b}.
\end{equation}

In the beginning of the section the condition $f^{\prime \prime }(0)=0$ was
imposed, here we will assume $f^{\prime \prime }(0)\neq 0.$ Thus we consider

\begin{equation}
\frac{\partial u}{\partial t}=D\nabla ^{2}u+f(u),\;f(0)=0,\;u(0,t)=u(1,t)=0.
\end{equation}
The solution $u=0$ is a steady state solution, so expanding near it we set

\[
u=\sum_{m=1}\varepsilon ^{m}v_{m}(t^{\prime },t^{\prime \prime },x),\;\frac{%
\partial }{\partial t}=\frac{\partial }{\partial t^{\prime }}+\varepsilon
\frac{\partial }{\partial t^{\prime \prime }},\;f^{\prime }(0)=\lambda
_{0}+\varepsilon \lambda _{1}.
\]
Substituting in Eq. (14) and equating terms $O(\varepsilon ),$ one gets

\[
v_{1}(t^{\prime },t^{\prime \prime },x)=\sum_{l=1}a_{l}(t^{\prime \prime
})\sin (\pi lx)\exp [(1-l^{2})\pi ^{2}].
\]
Let $\lambda _{0}=\pi $, and consider the equation $O(\varepsilon ^{2})$, we
set the secular term (independent of $t^{\prime }$) equal to zero, then

\begin{equation}
\frac{\mathrm{d}a_{1}}{\mathrm{d}t^{\prime \prime }}=\lambda
_{1}a_{1}+\left( \frac{b^{\prime }}{2}f^{\prime \prime }(0)\right)
a_{1}^{2},\;b^{\prime }=\frac{\int_{0}^{1}\sin ^{3}\pi x\;\mathrm{d}x}{%
\int_{0}^{1}\sin ^{2}\pi x\;\mathrm{d}x}.
\end{equation}
Thus the conditions for FAI are

\[
f^{\prime \prime }(0)>0,\;\lambda _{1}<0\;\mathrm{and}\;a_{1}(0)>-\frac{%
2\lambda _{1}}{b^{\prime }f^{\prime \prime }(0)}.
\]

For systems of two partial differential equations

\begin{equation}
\begin{array}{c}
\frac{\partial u_{1}}{\partial t}=D_{1}\nabla ^{2}u_{1}+f(u_{1},u_{2}),\;%
\frac{\partial u_{2}}{\partial t}=D_{2}\nabla ^{2}u_{2}+g(u_{1},u_{2}), \\
u_{1}(0,t)=\;u_{1}(1,t)\;=u_{2}(0,t)\;=u_{2}(1,t)=0, \\
f(0,0)=g(0,0)=0.
\end{array}
\end{equation}
Expanding near the steady state solutions $u_{1}=u_{2}=0,$ we get

\[
\begin{array}{c}
u_{1}=\sum_{m=1}\varepsilon ^{m}v_{m}(t^{\prime },t^{\prime \prime
},x),\;u_{2}=\sum_{m=1}\varepsilon ^{m}\omega _{m}(t^{\prime },t^{\prime
\prime },x), \\
\frac{\partial }{\partial t}=\frac{\partial }{\partial t^{\prime }}%
+\varepsilon \frac{\partial }{\partial t^{\prime \prime }}, \\
f_{1}=\frac{\partial f}{\partial u_{1}(0,0)}=\lambda _{11}+\varepsilon
\lambda _{12},\;f_{2}=\lambda _{21}+\varepsilon \lambda _{22}, \\
g_{1}=\mu _{11}+\varepsilon \mu _{12},\;g_{2}=\mu _{21}+\varepsilon \mu
_{22}.
\end{array}
\]
After some tedious calculations, we got the following conditions for FAI in
the system (16):

\begin{enumerate}
\item[(i)]  $(\pi ^{2}D_{1}-\lambda _{11})(\pi ^{2}D_{2}-\mu _{21})-\lambda
_{21}\mu _{11}=0.$

\item[(ii)]  $\lambda _{12}+\kappa \lambda _{22}=\;\frac{\mu _{12}}{\kappa }%
+\mu _{22}<0,\;$where$\;\kappa =\frac{\pi ^{2}D_{1}-\lambda _{11}}{\lambda
_{21}}.$

\item[(iii)]  $\;\frac{f_{11}}{2}+\kappa f_{12}+\frac{\kappa ^{2}}{2}f_{22}=%
\frac{g_{11}}{2\kappa }+g_{12}+\frac{\kappa ^{.}}{2}g_{22}>0.$

\item[(iv)]  $a_{1}(0)>-\frac{\lambda _{12}+\kappa \lambda _{22}}{b^{\prime
}\left( \frac{f_{11}}{2}+\kappa f_{12}+\frac{\kappa
^{2}}{2}f_{22}\right) },$\\
\end{enumerate}

where $b^{\prime }$ is defined in Eq. (15). Applying the condition (i) to
the spatial AHD, one gets

\[
(\pi ^{2}D_{1}+v_{1})(\pi ^{2}D_{2}+v_{2})<0,
\]
which is not possible, thus we find:\newline
\newline
\textbf{Proposition (3):} There is no finite amplitude instability for the
solution $p=1,\;q=0$ of the AHD game.

\section{Asymmetric prisoner's dilemma game}

In the prisoner's dilemma game, the possible strategies are cooperate (C)
and defect (D). The payoff matrices $A,\;B$ in Eq. (1) are given as follows:

\begin{equation}
A=\left[
\begin{array}{cc}
R_{1} & S_{1} \\
T_{1} & P_{1}
\end{array}
\right] ,\;B=\left[
\begin{array}{cc}
R_{2} & S_{2} \\
T_{2} & P_{2}
\end{array}
\right]
\end{equation}
such that \ $2R_{i}>T_{i}+S_{i},\ $and $T_{i}>R_{i}>P_{i}>S_{i}$, where $%
i=1,\;2$. The dynamical equations for the \ spatial asymmetric prisoner's
dilemma game (spatial APD) are:

\begin{equation}
\begin{array}{c}
\frac{\partial u_{1}}{\partial t}=D_{1}\nabla ^{2}u_{1}+ \\
u_{1}(1-u_{1})[-(P_{1}-S_{1})+u_{2}(P_{1}-S_{1}-T_{1}+R_{1})], \\
\frac{\partial u_{2}}{\partial t}=D_{2}\nabla ^{2}u_{2}+ \\
u_{2}(1-u_{2})[-(P_{2}-S_{2})+u_{1}(P_{2}-S_{2}-T_{2}+R_{2})],
\end{array}
\end{equation}
where $u_{1}(u_{2})$ is the fraction of cooperators in the first (second)
players population. The solution $u_{1}=u_{2}=0\;$is linearly asymptotically
stable. It represents the always defect strategy.

Two questions arise the first is: can diffusion stabilize the cooperation
solution \ $u_{1}=u_{2}=1$? And does the always defect solution have (FAI)
nonlinear instability? Using the techniques of the previous section we get:%
\newline
\newline
\textbf{Proposition (4):}

\begin{enumerate}
\item[(i)]  If $D_{i}\pi ^{2}>T_{i}-R_{i}$, then the cooperation solution is
linearly asymptotically stable.

\item[(ii)]  The always defect solution does not have FAI.
\end{enumerate}

\section{Telegraph reaction diffusion in spatial asymmetric games}

The standard spatial games depend on the familiar reaction-diffusion
equation
\begin{equation}
\frac{\partial u(x,t)}{\partial t}=D\frac{\partial ^{2}u(x,t)}{\partial x^{2}%
}+f(u).
\end{equation}
A basic weakness in this equation is that the flux $j$ reacts simultaneously
to the gradient of $u$ consequently an unbounded propagation speed is
allowed. This manifests itself in many solutions to Eq. (1) e.g. (if $f=0$),
then

\[
u(x,t)=\frac{1}{\sqrt{4\pi Dt}}e^{\frac{-x^{2}}{4Dt}},\;u(x,0)=\delta (x)\;%
\mathrm{i.e}\;u(x,t)>0\forall x,\;\forall t>0.
\]
This is unrealistic specially in biological and economical systems, where it
is known that propagation speeds are typically small. \ To rectify this
weakness, Fick's law is replaced by
\[
j+\tau \frac{\partial j}{\partial t}=D\frac{\partial u}{\partial x},
\]
and the resulting telegraph diffusion equation becomes
\[
\tau \frac{\partial ^{2}u}{\partial t^{2}}+\frac{\partial u}{\partial t}=D%
\frac{\partial ^{2}u}{\partial x^{2}}.
\]
The corresponding telegraph reaction diffusion (TRD) \ is\

\begin{equation}
\tau \frac{\partial ^{2}u}{\partial t^{2}}+(1-\tau \frac{\mathrm{d}f}{%
\mathrm{d}u})\frac{\partial u}{\partial t}=D\nabla ^{2}u+f(u)
\end{equation}
The time constant $\tau $ can be related to the memory effect of the flux $j$
as a function of \ the distribution $u$ as follows: Assume that [Compte and
Metzle 1997]

\begin{equation}
\ j(x,t)=-\int_{0}^{t}K(t-t^{\prime })\frac{\partial u(x,t^{\prime })}{%
\partial x}\;\mathrm{d}t^{\prime },
\end{equation}
hence

\[
j+\tau \frac{\partial j}{\partial t}=-\tau K(0)u(x,t)\;-\int_{0}^{t}\left(
\tau \frac{\partial K(t-t^{\prime })}{\partial t}+K(t-t^{\prime })\right)
\frac{\partial u}{\partial x}\;\mathrm{d}t^{\prime }.
\]
This equation is equivalent to the telegraph equation if
\[
K(t)=\frac{D}{\tau}\exp (\frac{-t}{\tau}).
\]
This lends further support that
TRD is more suitable for economic and biological systems than the
ordinary diffusion equation since e.g. it is known that we take
our decisions according to our previous experiences, so memory
effects are quite relevant. Further evidence comes from the work
of Chopard and Droz [Chopard and Droz 1991], where they have shown
that starting from discrete time and space then the continuum
limit does not give the standard reaction diffusion but the
telegraph one.

Since it is known that new technologies, habits etc... takes time to spread,
we believe that TRD equation is more relevant than the ordinary diffusion
equation in modelling economic and biological systems [Ahmed et al 2001].

Now we apply TRD to spatial APD game. The TRD for a system of equations are
[Hadeler 1998]

\begin{equation}
\begin{array}{c}
\tau \frac{\partial ^{2}u_{i}}{\partial t^{2}}+\frac{\partial u_{i}}{%
\partial t}-\tau \sum_{j}\frac{\partial u_{j}}{\partial t}\frac{\partial
f_{i}}{\partial u_{j}}\;= \\
D\nabla ^{2}u_{i}+f_{i}(u_{1},u_{2},\ldots ,u_{n}),
\end{array}
\end{equation}
hence applying it to the APD (18), we get

\begin{equation}
\begin{array}{c}
\tau \frac{\partial ^{2}u_{1}}{\partial t^{2}}+\frac{\partial u_{1}}{%
\partial t}-\tau \frac{\partial u_{1}}{\partial t}(1-2u_{1})[-(P_{1}-S_{1})+
\\
u_{2}(P_{1}-S_{1}-T_{1}+R_{1})]=D_{1}\nabla ^{2}u_{1}+ \\
u_{1}(1-u_{1})[-(P_{1}-S_{1})+u_{2}(P_{1}-S_{1}-T_{1}+R_{1})], \\
\tau \frac{\partial ^{2}u_{2}}{\partial t^{2}}+\frac{\partial u_{2}}{%
\partial t}-\tau \frac{\partial u_{2}}{\partial t}(1-2u_{2})[-(P_{2}-S_{2})+
\\
u_{1}(P_{2}-S_{2}-T_{2}+R_{2})]=D_{2}\nabla ^{2}u_{2}+ \\
u_{2}(1-u_{2})[-(P_{2}-S_{2})+u_{1}(P_{2}-S_{2}-T_{2}+R_{2})].
\end{array}
\end{equation}

The following question arises: Can diffusion stabilize the cooperation
solution \ $u_{1}=u_{2}=1$? Using the techniques of the second section, we
get\newline
\newline
\textbf{Proposition (5):} If the following conditions are satisfied

\begin{equation}
\begin{array}{c}
D_{i}\pi ^{2}>T_{i}-R_{i},\;1>\tau (T_{i}-R_{i}), \\
4\tau (D_{i}\pi ^{2}-T_{i}+R_{i})\leq [-1+\tau (T_{i}-R_{i})]^{2},\;i=1,2,
\end{array}
\end{equation}
then the cooperation solution is linearly asymptotically stable.\newline
\newline
\textbf{Proof. }Assume that

\[
u_{1}=1-\varepsilon _{1}\exp (\sigma _{1}t)\sin (\pi
x),\;u_{2}=1-\varepsilon _{2}\exp (\sigma _{2}t)\sin (\pi x).
\]
Substituting one gets

\[
\begin{array}{c}
\sigma _{1}=\frac{1}{2\tau }[\left( -1+\tau (T_{1}-R_{1})\right) \pm \\
\sqrt{\left( -1+\tau (T_{1}-R_{1})\right) ^{2}-4\tau \left(
R_{1}-T_{1}+D_{1}\pi ^{2}\right) }], \\
\sigma _{2}=\frac{1}{2\tau }[\left( -1+\tau (T_{2}-R_{2})\right) \pm \\
\sqrt{\left( -1+\tau (T_{2}-R_{2})\right) ^{2}-4\tau \left(
R_{2}-T_{2}+D_{2}\pi ^{2}\right) }].
\end{array}
\]
Stability requires that the real part of $\sigma _{i},\;i=1,2$ is negative.
The first two conditions in the proposition guarantee this requirement.
Furthermore since $u_{i},\;i=1,2$ are real and nonnegative, then the term
under the square root should be nonnegative. The third condition of the
proposition guarantees $u_{i}\geq 0$.\newline
\newline
Thus the conditions for cooperation stability for TRD are more stringent
than those for ordinary diffusion (c.f. proposition (4)).

\section{\ Asymmetric game of parental investment}

Parents are faced with the decision whether to care for the offsprings or to
desert. A model has been given for this asymmetric game [Krebs and Davies
2000]. Let $p_{0},\;p_{1},\;p_{2},$ be the probabilities of survival of
offsprings which are not cared for, cared for by a single parent and cared
for by both parents, respectively, then $\;p_{0}<p_{1}<p_{2}.$ A deserting
male has a chance $q$ of mating again while a caring (deserting) female has $%
w_{1}(w_{2})$ offsprings. The payoff matrices for male (female)
corresponding to the strategies C (care) or D (desert) are denoted by $A(B),$
and given by
\begin{equation}
A=\left[
\begin{array}{ll}
w_{1}p_{2} & w_{2}p_{1} \\
w_{1}p_{1}(1+q) & w_{2}p_{0}(1+q)
\end{array}
\right] ,\;\;B=\left[
\begin{array}{ll}
w_{1}p_{2} & w_{1}p_{1} \\
w_{2}p_{1} & w_{2}p_{0}
\end{array}
\right] .
\end{equation}
The spatial asymmetric equations for the above game are:

\[
\begin{array}{c}
\frac{\partial u}{\partial t}%
=u[w_{1}p_{2}v+w_{2}p_{1}(1-v)-w_{1}p_{2}uv-w_{2}p_{1}u(1-v)- \\
w_{1}p_{1}(1+q)v(1-u)-w_{2}p_{0}(1+q)(1-u)(1-v)]+D_{1}\frac{\partial ^{2}u}{%
\partial x^{2}}+D_{12}\frac{\partial ^{2}v}{\partial x^{2}}, \\
\frac{\partial v}{\partial t}%
=v[w_{1}p_{2}u+w_{1}p_{2}(1-u)-w_{1}p_{2}uv-w_{2}p_{1}u(1-v)- \\
w_{1}p_{1}v(1-u)-w_{2}p_{0}(1-u)(1-v)]+D_{2}\frac{\partial ^{2}v}{\partial
x^{2}}+D_{21}\frac{\partial ^{2}u}{\partial x^{2}}.
\end{array}
\]
In this system, we introduced both ordinary and cross diffusion. Cross
diffusion is the diffusion of one type of species due to the presence of
another [Okubo 1980]. This phenomena is abundant in nature e.g.
predator-prey systems where the predator diffuses towards the regions where
the prey is more abundant. On the other hand the prey tries to avoid
predators by diffusing away from it. Another area of application is in
epidemics where susceptible individuals try to avoid infected ones.

Here we will see that ordinary diffusion is unable of destabilizing the
diffusionless ESS:

\begin{enumerate}
\item[(i)]  ESS1 where both male and female desert i.e. $(u=0,v=0)$. It
requires $w_{2}p_{0}>w_{1}p_{1}\;$and$\;p_{0}(1+q)>p_{1}.$

\item[(ii)]  ESS2 where male cares and female desert i.e. $(u=1,v=0)$. It
requires $w_{2}p_{1}>w_{1}p_{2}\;$and$\;p_{0}(1+q)<p_{1}.$

\item[(iii)]  ESS3 where female cares and male desert i.e. $(u=0,v=1)$. It
requires $w_{1}p_{1}>w_{2}p_{0}\;$and$\;p_{1}(1+q)>p_{2}.$

\item[(iv)]  ESS4 where both male and female care i.e. $(u=1,v=1)$. It
requires $w_{1}p_{2}>w_{2}p_{1}\;$and$\;p_{1}(1+q)<p_{2}.$
\end{enumerate}

Following steps similar to the previous games, we get\newline
\newline
\textbf{Proposition (6):} The solution ESSi, $i=1,2,3,4$ is destabilized if
the following condition is satisfied:
\begin{equation}
D_{12}D_{21}\pi ^{4}>(D_{1}\pi ^{2}-a_{i})(D_{2}\pi ^{2}-b_{i})
\end{equation}
where
\begin{eqnarray*}
a_{1} &=&w_{2}p_{1}-w_{2}p_{0}(1+q),\;\;b_{1}=w_{1}p_{1}-w_{2}p_{0}, \\
a_{2} &=&-w_{2}p_{1}+w_{2}p_{0}(1+q),\;\;b_{1}=w_{1}p_{2}-w_{2}p_{1}, \\
a_{3} &=&w_{1}p_{2}-w_{1}p_{1}(1+q),\;\;b_{1}=-w_{1}p_{1}+w_{2}p_{0}, \\
a_{4} &=&-w_{1}p_{2}+w_{1}p_{1}(1+q),\;\;b_{1}=w_{2}p_{1}-w_{1}p_{2}.
\end{eqnarray*}
\newline
\newline
Notice that all $a_{i},\;b_{i},\;i=1,2,3,4$ are negative hence ordinary
diffusion cannot destabilize the ESS.

Applying the above procedure to the battle of the sexes [Schuster and
Sigmund 1981] where the female has two strategies coy or fast while the male
can be either faithful or philanderer. The male (female) payoff matrix is $%
A(B)$
\[
A=\left[
\begin{array}{ll}
0 & -10 \\
-2 & 0
\end{array}
\right] ,\;\;B=\left[
\begin{array}{ll}
0 & 5 \\
3 & 0
\end{array}
\right] ,
\]
hence the spatial battle of the sexes equations are

\begin{equation}
\begin{array}{c}
\frac{\partial u}{\partial t}=u(1-u)(-10+12v)+D_{1}\frac{\partial ^{2}u}{%
\partial x^{2}}, \\
\frac{\partial v}{\partial t}=v(1-v)(5-8u)+D_{2}\frac{\partial ^{2}v}{%
\partial x^{2}}.
\end{array}
\end{equation}
\newline
\newline
\textbf{Proposition (7): }Diffusion stabilizes the internal equilibrium of
the system (27).\newline
\newline
\textbf{Proof. }There is a unique internal homogeneous equilibrium solution $%
E=(5/8,5/6)$, which (for the diffusionless case) is stable but not
asymptotically stable. Substituting with
\[
u=\frac{5}{8}+\varepsilon \exp (\sigma t)\sin \pi x,\;\;v=\frac{5}{6}%
+\varsigma \exp (\sigma t)\sin \pi x,\;1\geq x\geq 0,
\]
in Eq. (27), and linearizing in $\varepsilon,\;\zeta $, one gets
\[
(\sigma +D_{1}\pi ^{2})\varepsilon =\frac{45}{16}\zeta ,\;(\sigma +D_{2}\pi
^{2})\zeta =-\frac{10}{9}\varepsilon .
\]
Hence
\[
[(\sigma +D_{1}\pi ^{2})(\sigma +D_{2}\pi ^{2})+\frac {25}{8}]\zeta =0,
\]
i.e.
\[
(\sigma +D_{1}\pi ^{2})(\sigma +D_{2}\pi ^{2})+\frac {25}{8}=0,
\]
then
\[
\sigma ^{2}+\sigma (D_{1}\pi ^{2}+D_{2}\pi ^{2})+(\frac
{25}{8}+D_{1}D_{2}\pi ^{4})=0.
\]
Therefore the real part of $\sigma $ is negative, then the internal
equilibrium (in the presence of diffusion) is asymptotically stable. It is
clear that if the diffusion coefficients are set equal to zero $%
(D_{1}=D_{2}=0)$, then one regains the stability but not asymptotic
stability. This completes the proof.

\section{Conclusions}

Based on replicator equations, a mathematical approach for the analysis of
spatial asymmetric games is introduced. Some questions regarding spatial
stability for asymmetric hawk-dove and the asymmetric prisoner's dilemma
(APD) games are answered. Telegraph reaction diffusion equation is applied
for stability analysis of the asymmetric prisoner's dilemma game. Asymmetric
game of parental investment is studied in the presence of both ordinary and
cross diffusions. Ordinary diffusion cannot destabilize the ESS for this
game. Conditions for destabilizing the ESS are given in the case of cross
diffusion.\newline
\newline
\textbf{Acknowledgments}\newline

We thank the referees for their helpful comments.\newline
\newline
\textbf{References}

\begin{enumerate}
\item[ ]  Ahmed E., Abdusalam H. A. and Fahmy I. (2001), Telegraph reaction
diffusion equations. Int. J. Mod. Phys. C 12, 717-726.

\item[ ]  Boerlijst M. C. and Hogeweg P. (1995), Attractors and spatial
patterns in hypercycles with negative interactions. J. Theor. Biol. 176,
199-210.

\item[ ]  Chopard B. and Droz M. (1991), Cellular automata model for the
diffusion equation, J. Stat. Phys. 64, 859-892.

\item[ ]  Compte A. and Metzle R. (1997), The generalized Cattaneo equation
for the description\ of anomalous transport processes. J. Phys. A 30,
7277-7289.

\item[ ]  Cressman R. and Vickers G. T. (1997), Spatial and density effects
in evolutionary game theory. J. Theor. Biol. 184, 359-369.

\item[ ]  Hadeler K. P. (1998). in O. Diekman et al (eds), Mathematics
inspired by Biology. Springer, Berlin.

\item[ ]  Hofbauer J., Hutson V. and Vickers G. T. (1997), Travelling waves
for games in economics and biology. Nonlinear Analysis, Theory. Methods and
Applications 30, 1235-1244.

\item[ ]  {Hofbauer J. and Sigmund K. (1998), Evolutionary games and
population dynamics.\ Cambridge University Press, Cambridge.}

\item[ ]  Krebs J. R. and Davies N. B. (2000), An introduction to behavioral
ecology. Blakwell scientific pub., Oxford.

\item[ ]  Okubo A. (1980), Diffusion and ecological problems. Springer,
Berlin.

\item[ ]  Schuster P. and Sigmund K. (1981), Coyness philandering and stable
strategies. Anim. Behav. 29, 186-192.

\item[ ]  Stuart A. (1989), Nonlinear instability in dissipative finite
difference scheme.\ Siam Rev. 31, 191-220.
\end{enumerate}

\end{document}